\documentclass{PoS}

      \newcommand{\tr} { \mathrm{Tr}}

\title{Analytic continuation of finite density QCD with heavy quarks in the strong coupling region}

\ShortTitle{Analytic continuation of finite density QCD with heavy quarks}
\author{Shinji Ejiri\\
       Graduate School of Science and Technology, Niigata University, Niigata 950-2181, Japan\\
        E-mail: \email{ejiri@muse.sc.niigata-u.ac.jp}}
\author{\speaker{Hiroshi Yoneyama}
\\
       Department of Physics, Saga University, 840-8502 Saga, Japan\\
        E-mail: \email{yoneyama@cc.saga-u.ac.jp}}

\abstract{Complex nature of finite  density QCD with heavy quarks in the strong coupling region is studied. 
For this purpose, we consider the effective potential as a function of Polyakov line, 
and study   
  thermodynamic singularities and associated Stokes boundaries  in the complex chemical potential ($\mu$) plane. 
  We also perform an explicit analytic continuation of the first order transition  and crossover lines  in the complex $\mu$ plane.
}

\FullConference{The 33rd International Symposium on Lattice Field Theory\\
		14 -18 July 2015\\
		Kobe International Conference Center, Kobe, Japan*}

\begin{document}

\section{Introduction}
In this talk, we discuss   the phase structure in the  quark complex  chemical potential ($\mu$)  plane of the Polyakov line model, which is  an effective theory of finite density QCD with heavy quarks in the strong coupling region.  The purpose of  studying   in the complex $\mu$ plane is twofold. One is that  the Lee-Yang zeros~\cite{YL,LY,IPZ}, the edge singularities and the Stokes boundaries  are deeply associated with     the   phase structures,  and  may provide  useful information of    the QCD critical point~\cite{S, EY, ESY}.  The other concerns   the validity of the imaginary chemical potential method used in the Monte Carlo simulations~\cite{Lo, deFP, deFP2, DL, DL2, CCDP, WLC,  CCEMP,NN2, CCEPS}, which relies on the  analytic continuation  in the complex $\mu$ plane.  
 \par
In the space of parameters associated with   the hopping parameter, the gauge coupling and $\mu$, this model reveals an interesting structure involving first order phase transition lines, crossovers and critical endpoints.   
  To discuss their  complex nature,   we calculate    the effective potential as a function of the Polyakov line,   adopting    a variational method based on    the Legendre transformation in the mean field framework. 
 By extending $\mu$ to the  complex plane, we investigate thermodynamic singularities,   the edge singularities and the associated Stokes boundaries.  
 Our study  may   hopefully provide a useful  view of  the expected QCD critical point.
\par
In the following section, the effective potential is calculated.  In Sect. 3, we study the model in the complex $\mu$ plane. Critical endpoint as a singularity, the edge singularity, crossover and Stokes boundaries are discussed.  The first order phase transition line is analytically continued to   the complex plane.    Conclusions are given in Sect. 4.  
\section{Effective potential}
\begin{figure}
\vspace{-2mm}
        \centerline{\includegraphics[width=6cm, height=5
cm]{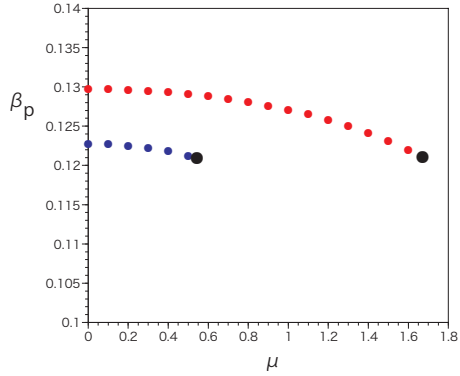}}
\caption{Phase diagram in $\mu-\beta_p$ plane. First order phase transition lines are shown  for $\kappa=$0.02 (red) and 0.05 (blue). Black filled symbol  indicates  the critical endpoints. }
\label{fig:phdg-k02}
\end{figure}

The Polyakov line  model, which  is an effective theory of finite temperature and finite density QCD in the strong coupling region  at the leading order in the hopping  parameter expansion,  
is defined  by
\begin{eqnarray}
Z&=&\int [dU] e^{-S}, \label{eq:pf} \\
S&=&- \beta_p \sum_{x,x'}\sum_{i=1}^{d}\mathrm{Tr} \ U_x \mathrm{Tr} \ U_{x'}^\dagger +c.c. -\kappa \sum_{x} \left(e^{\mu} \mathrm{Tr}\  U_{x}+e^{-\mu} \mathrm{Tr}\  U_{x}^\dagger \right), 
\label{model}
\end{eqnarray}
where  $\beta_p$ and  $\kappa$ are two parameters defining the model, and the   $\mu$ is chemical potential.  The space dimensionality  $d$ is fixed to 3 throughout  the calculations.
In this report  we use the mean field theory  based on the variational method in terms of the Legendre transformation by
introducing  two variational parameters $K$ and $\hat{K}$ to the partition function\footnote{Greensite and Splittorff~\cite{GS} has investigated the model using the mean field method  in the different framework from ours  and found that its  phase diagram agrees  fairly well  with that of Monte Carlo simulations~\cite{MG}. } ; 
\begin{eqnarray}
Z=\int & [dU] &  \exp\left[   \beta_p \sum_{x,x'}\sum_{i=1}^{d}\mathrm{Tr} \ U_x \mathrm{Tr} \ U_{x'}^\dagger +c.c. - K \sum_x \tr U_x -\hat{K}\sum \tr U^\dagger_x\right] \nonumber \\
& \times &\exp\left[ (h+K)\tr U_x +(\hat{h}+\hat{K}) \tr U^\dagger_x \right],
\end{eqnarray}
where $h=\kappa e^{\mu}$ and $\hat{h}=\kappa e^{-\mu}$. 
  Defining  a  partition function characterize by $K$ and $\hat{K}$ with the site-independent measure on a $N_s$-site lattice
\begin{eqnarray}
Z_{K,\hat{K}}\equiv \left(z_{K,\hat{K}}\right)^{N_s}, \quad z_{K,\hat{K}}=\int [dU] e^{ (h+K)\tr U +(\hat{h}+\hat{K}) \tr U^\dagger}, 
\label{zk}
\end{eqnarray}
$Z$ is interpreted as  an expectation value  
\begin{eqnarray}
Z=\langle \exp\left[   \beta_p \sum_{x,x'}\sum_{i=1}^{d}\mathrm{Tr} \ U_x \mathrm{Tr} \ U_{x'}^\dagger +c.c. - K \sum_x \tr U_x -\hat{K}\sum \tr U^\dagger_x\right]  \rangle_{K,\hat{K}}  Z_{K,\hat{K}},
\label{expect}
\end{eqnarray}
where  the expectation value $\langle \cdot \rangle_{K,\hat{K}}$ specified by $K$ and $\hat{K}$ is taken with respect to  Eq.~(\ref{zk}). 
Application  of the  Jensen's inequality $\langle e^{A} \rangle \geq e^{\langle A \rangle}$
 to Eq.~(\ref{expect})   leads to 
an  inequality concerning the  free energy density $f=-(1/N_s)\log Z$
\begin{eqnarray}
f \leq  f_{K, \hat{K}}  \label{frengy},
\end{eqnarray}
 where $ f_{K, \hat{K}}$  is an approximate free energy density 
 \begin{eqnarray}
 f_{K, \hat{K}}\equiv  - \log z_{K,\hat{K}} 
 - \beta_p d \ \left( \langle \tr U\rangle_{K,\hat{K}}  \langle \tr U^{\dagger}\rangle_{K,\hat{K}}+ c.c. \right)+K \langle \tr U\rangle_{K,\hat{K}}+\hat{K}  \langle \tr U^{\dagger} \rangle_{K,\hat{K}}.
\label{ineq} 
\end{eqnarray}
The parameters $K$ and $\hat{K}$ are fixed   so as to minimize $ f_{K, \hat{K}}$ as the best approximation,  
and consequently, we have
\begin{eqnarray}
 f_{\bar{K}, \bar{\hat{K}}}= - \log z_{\bar{K}, \bar{\hat{K}}} +\frac{1}{2\beta_p d}\bar{K} \bar{\hat{K}},
\end{eqnarray}
where 
\begin{eqnarray}
\bar{K}=2\beta_p d \  \langle \tr U^{\dagger} \rangle_{\bar{K},  \bar{\hat{K}}} , \quad \bar{\hat{K}}=2\beta_p d \   \langle \tr U \rangle_{\bar{K},  \bar{\hat{K}}}.
\label{KKBar}
\end{eqnarray}
 In order to have the effective potential $\Omega(M)$ as a function of expectation value of $M=\langle \tr \ U \rangle$
 through  the Legendre transform of $ f_{\bar{K}, \bar{\hat{K}}}$,  
we introduce  an  ``external field" $h_e$    conjugate to $M$ so that
\begin{eqnarray}
\frac{\partial  f_{\bar{K}, \bar{\hat{K}}}}{\partial h_e}=-M.
\label{Leg1}
\end{eqnarray}
Inverting  Eq.~(\ref{Leg1})  to obtain $h_e(M)$, $ f_{\bar{K}, \bar{\hat{K}}}$ is Legendre-tranformed to $\Omega(M)$
\begin{eqnarray}
\Omega(M)= f_{\bar{K}, 2M\beta_pd }(h_e(M))+h_e(M)M,
\label{Leg2}
\end{eqnarray}
where 
\begin{eqnarray}
\frac{\partial \Omega(M)}{\partial M}=h_e. 
\label{he}
\end{eqnarray}
\begin{figure}
\vspace{-5mm}
        \centerline{\includegraphics[width=6cm, height=5
cm]{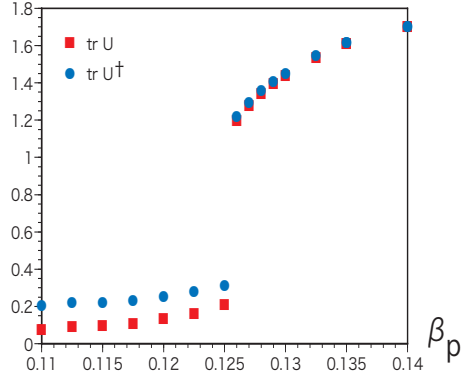}}
\caption{
Expectation values of $\tr \ U$ and $\tr \ U^{\dagger}$  as a function of $\beta_p$.   $\kappa=0.02$ and $\mu=1.2$.  }
\label{fig:expvalue-MMdag}
\end{figure}
Fig.~\ref{fig:expvalue-MMdag} indicates,  as a function of $\beta_p$,  the expectation values of $\tr \ U$ and $\tr \ U^{\dagger}$, the former is identified as a global minimum of $\Omega(M)$ satisfying $h_e=0$ in Eq.~(\ref{he}), and the latter   from the corresponding $\bar{\hat{K}}/(2\beta_p d )$.  It is seen that  a first order phase transition occurs    at  $\beta_p=0.12579$.   In Fig.~\ref{fig:phdg-k02}, the first order phase transition lines and the critical endpoints   are  shown for $\kappa=0.02$ and 0.05  in $\mu-\beta_p$ plane.   As $\kappa$ goes up, the first order phase transition line shrinks  and disappears  at $(\mu,\beta_p) \approx (0, 0.12068)$ for  $\kappa_c\approx 0.05905$.  This phase structure  is in agreement with that in \cite{GS}.   

\section{Complex $\mu$ plane}
\subsection{Singularities}
In this section, we discuss the nature in the complex $\mu$ plane.
The singularity occurs when the global minimum of $\Omega(M)$ becomes unstable, i.e., the followings are  satisfied  
\begin{eqnarray}
\frac{\partial \Omega(M)}{\partial M}=0, \quad \frac{\partial^2 \Omega(M)}{\partial M^2}=0,
\end{eqnarray}
which are  equivalent to 
\begin{eqnarray}
h_e(M)=0,  \quad \frac{\partial h_e(M)}{\partial M}=0
\label{hhm}
\end{eqnarray}
in our formulation.
For the equality on the right hand side in  Eq.~(\ref{hhm}), we use  an explicit expression for $\partial h_e/ \partial M$ 
\begin{eqnarray}
 \frac{\partial h_e}{\partial M}  =  \frac{1}{\langle \left(\tr \ U\right)^2\rangle_c}\left[
\left(
2\beta_pd \langle \ \tr \ U \  \tr \ U^\dagger\rangle_c -1
\right)^2 -(2\beta_pd)^2\langle \left(\tr \ U\right)^2\rangle_c \langle ( \tr \ U^\dagger )^2\rangle_c
\right],
\label{hem}
\end{eqnarray}
where
 \begin{eqnarray}
\langle AB \rangle_c\equiv \langle AB \rangle_{\bar{K},2M\beta_pd} -\langle A \rangle_{\bar{K},2M\beta_pd}\langle B \rangle_{\bar{K},2M\beta_pd}. 
\end{eqnarray}
 \par
\begin{figure}
\vspace{-0mm}
        \centerline{\includegraphics[width=6cm, height=5
cm]{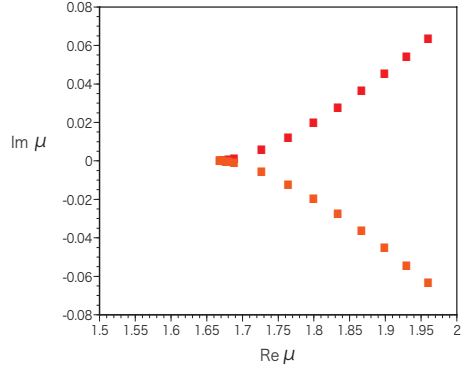}}
\caption{
Locations of the singular points for $\kappa=0.02$ in the complex $\mu$ plane.  The critical point is located at the left edge of the plots ($\mu=1.6709 \equiv \mu_E$).   
    }
\label{fig:singularity-k02}
\end{figure}

Fig.~\ref{fig:singularity-k02} indicates the locations of   the singular points as a solution to Eq.~(\ref{hhm}) in the complex $\mu$ plane for $\kappa=0.02$.  
 The left edge of the plots at $\mu=1.6709 \equiv \mu_E$ ($\beta_p=0.12122$)
  corresponds to the endpoint of the aforementioned  first order phase transition line in Fig.~\ref{fig:phdg-k02}. 
 For Re~$\mu > \mu_E$,   a  complex conjugate pair appears as the edge singularities  corresponding  to different value of $\beta_p$  ($0.117 \leq \beta_p \leq 0.12122$). 
  A fit to  the upper half part of the singular points in Fig.~\ref{fig:phdg-k02} give     Im~$\mu \propto$  Re~$(\mu-\mu_E)^{3/2}$.  \par
 Moving away of the singularities from the real axis  for Re~$\mu > \mu_E$ causes a crossover phenomenon on the real $\mu$  axis.    
In agreement with what  was shown in \cite{ESY}, it turns out  that  the location of the crossover, which is measured as a peak of the  susceptibility $\chi$ associated with the Polyakov loop
 \begin{eqnarray}
\chi\equiv \left. 1/\left(\frac{\partial^2 \Omega(M) }{\partial M^2}\right)\right|_{\bar{M}},
\end{eqnarray}
    agrees with   the real part of the location of the singularity in the complex $\mu$ plane.  
\subsection{Stokes boundaries}

\begin{figure}
\vspace{-0mm}
        \centerline{\includegraphics[width=6cm, height=4
cm]{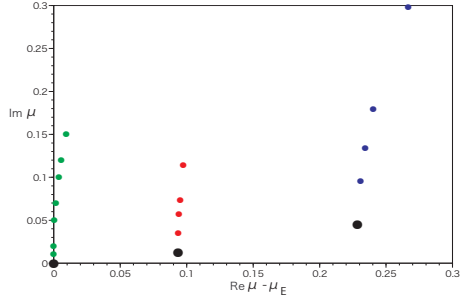}}
\caption{Behaviors of the Stokes boundaries for $\kappa=0.02$.  Filled circles (black) are singular points for three values of $\beta_p$, $0.12123  (=\beta_E)$, 0.12 and 0.118,  and   the Stokes boundary emanates from each of them.  }
\label{fig:stokesline-k02}
\end{figure}

For each singular point,  an associated cut, the Stokes boundary,  is connected to it  in the complex $\mu$ plane.  To locate where it is, we perform the analytic continuation of the global minimum in the complex $\mu$ plane.  For this,   setting $\mu-\mu_E=\rho e^{i\theta}$ and varying $\theta$ from $0$ to $\pi$ for every  fixed value of $\rho$, we trace the movement of the global minimum  as was done  in \cite{ESY}.      In the course of variation, a jump of the global minimum occurs at a point where the real part of the potential  Re~$\Omega(M)$ satisfies the following 
\begin{eqnarray}
{\rm Re}~\Omega(M_1)={\rm Re}~\Omega(M_2),
\label{ReV12}
\end{eqnarray}
where $M_1$ and $M_2$ are the location of the  two different  global minima analytically continued from those  on the real axis. 
 In Fig.~\ref{fig:stokesline-k02},  the behaviors of the Stokes boundaries  in the neighborhood of $\mu_E$  are shown.   For $\beta_p=0.12123$,  the Stokes boundary goes upright ($\varphi=\pi/2$) from the origin ($\mu=\mu_E$).   As $\beta_p$ deviates from $\beta_E$,  the singularity moves away from the real axis,  and corresponding Stokes boundary emanates from it with slightly   increasing  $\varphi $.   
\subsection{Analytic continuation of the first order phase transition line}
\begin{figure}
\vspace{-0mm}
        \centerline{\includegraphics[width=6cm, height=6
cm]{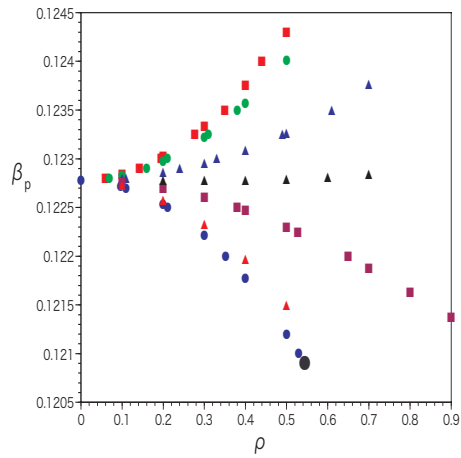}}
\caption{Analytic continuation of the first order phase transition line as a function of $\rho$  in the vicinity of the origin, where $\mu=\rho e^{i\theta}$.   $\kappa=0.05$.  Seven values of $\theta$ are  chosen;   $0, \pi/10, 2\pi/10, \pi/4, 3\pi/10, 4\pi/10$ and $\pi/2$ (from bottom to top).      The black filled symbol  for $\theta=0$  indicates the critical endpoint on the real $\mu$ axis. }
\label{fig:AC-firstorder}
\end{figure}

In dealing with the sign problem in  lattice simulations of finite density QCD, the imaginary chemical potential  method  relies on  the validity of   the analytic continuation from the real $\mu$ axis to the imaginary one.  In order  
 to   explicitly perform the analytic continuation, we study 
 how the first order phase transition line  on the real $\mu$ axis  persist in  the complex plane.  For this purpose,  we trace the global minimum of $\Omega(M)$ in the complex plane as done in the previous subsection. 
With the parametrization $\mu=\rho e^{i\theta}$, we vary the value of  $\rho$ for a   fixed value of $\theta$, and search  a  location in the complex $\mu$ plane where a jump of the global minimum occurs.  Fig.~\ref{fig:AC-firstorder} indicates the   first order phase transition lines  in the  $\rho - \beta_p$ plane for several fixed $\theta$ values in the region of   $0\leq \theta \leq \pi/2$, where   $\kappa$ is fixed to $0.05$.   It is found that the first order phase transition line varies   smoothly from the real axis ($\theta=0$) to the imaginary one   ($\theta=\pi/2$).   Monotonically decreasing curve at $\theta=0$ (real $\mu$ axis) gradually turns into  increasing one as $\theta$ increases, and for $\theta=\pi/4$, it becomes  almost constant, suggesting a behavior of the first order phase transition surface as 
$\beta_p(\mu)\propto {\rm Re} \ \mu^2$.
\section{Remarks}
We  investigated  the phase structure  in the complex chemical potential  plane of finite  density QCD with heavy quarks in the strong coupling region by adopting the Polyakov line model. 
We considered the effective potential as a function of Polyakov line, and studied    
  thermodynamic singularities and associated Stokes boundaries  in the complex chemical potential plane.
  By explicitly  tracing  the global minima,  we also performed  an  analytic continuation of the first order transition   lines   and found that the transition line is smoothly continued from the real $\mu$ axis to the pure imaginary axis.  
   
   Monte Carlo calculations of the  effective potential as a function of the Polyakov line of QCD with heavy quarks was done in the literature~\cite{SEA}.   Its  extension to the complex $\mu$ and the comparison with the results presented here will 
 be presented elsewhere\cite{EY2}.   
\acknowledgments
This work is in part supported by JSPS KAKENHI Grant
No. 26400244, No. 26287040.

\end{document}